\documentstyle[aps,floats,prl,preprint,psfig]{revtex}
\newcommand{\be}{\begin{equation}}
\newcommand{\ee}{\end{equation}}
\newcommand{\ba}{\begin{eqnarray}}
\newcommand{\ea}{\end{eqnarray}}
\begin{document}
\title{High-pressure phase diagram of the exp-6 model: \\
The case of Xenon}
\author{Franz Saija$^1$~\cite{aff1} and Santi Prestipino$^2$~\cite{aff2}}
\address{$^1$ CNR, Istituto per i Processi Chimico-Fisici, sez. Messina, \\
via La Farina 237, 98123 Messina, Italy \\
$^2$ Universit\`a degli Studi di Messina, Dipartimento di Fisica, \\
and Istituto Nazionale per la Fisica della Materia, \\
Contrada Papardo, 98166 Messina, Italy}
\maketitle
\begin{abstract}
~~We investigated numerically the high-temperature/high-pressure phase
diagram of Xenon as modelled through the exp-6 interaction potential,
which is thought to provide a reliable description of the thermal
behaviour of rare gases under extreme conditions.
We performed a series of extensive $NVT$ Monte Carlo simulations which,
in conjunction with exact computation of the solid free energy
by the Frenkel-Ladd method, allowed us to precisely locate the freezing
and the melting thresholds at each temperature.
We find that, under isothermal compression, the exp-6 fluid freezes
directly into a FCC solid; however, above 4500 K, an intermediate BCC
phase becomes stable in a narrow range of pressures.
The chemical potential of the HCP phase does never significantly differ
from that of the FCC solid of equal $T$ and $P$, though the former is
found to slightly overcome the latter.
We discuss our results in the light of previous numerical studies of the
same model system and of the experimental data available for Xenon.
\end{abstract}
\thispagestyle{empty}

\vspace{2mm}

\vspace{2mm}

\newpage
%
%
\section{Introduction}
\setcounter{page}{1}

In the last decade, there has been an increasing interest in the
high-temperature/high-pressure (HT/HP) properties of many materials%
~\cite{Hemley,Scandolo,Shouten}, as realized for instance in the
deep core of the planets of our Solar system.
Such peculiar conditions do in fact provide a new means for testing
current condensed-matter theories, since squeezing matter to planetary
pressure and heating it up to a few thousand degrees can trigger various
forms of structural reorganization both at the macroscopic and at
the molecular level (see, for instance, the two paradigmatic cases
of water and methane), or even cause modifications in the electronic
transport properties (as it happens for {\it e.g.} hydrogen).
Since down-the-earth experiments are somehow limited (pressure can
hardly be pushed over a certain threshold and, more important, any
huge compression is plagued by severe non-hydrostaticity problems),
the only possible insights into the transformations undergone by
many substances at extreme conditions do often come from numerical
simulation -- provided, however, a theoretical model of that
substance is amenable to careful investigation and numerical
errors are kept under control.

Due to their closed electronic shells, (heavy) rare gases are usually
believed to be the simplest substances of all, hence it may appear
somewhat strange that their thermal behaviour in the HT/HP regime
({\it i.e.}, in the very-dense fluid and solid regions) is not yet
well assessed~\cite{Jephcoat1}.
Despite the fact that one can get rid of quantum-mechanical considerations
almost completely (at a temperature as high as 3500 K, it was shown in
Ref.\,\cite{Belonoshko1} that electronic excitations do not play any
significant role up to pressures of order 50 GPa), a classical
interaction potential that accurately reproduces the thermodynamic
properties of rare gases cannot be of the simple Lennard-Jones form.
In fact, while performing very well at ambient conditions, the
Lennard-Jones potential loses much of its reliability when temperature
and pressure take huge values.
For high system densities, three-body contributions to the effective
potential are likely to be as important as the two-body
term~\cite{Loubeyre}, with the effect that the atomic core is softer
than implied by the Lennard-Jones interaction.
If one insists in using a two-body effective potential, a more reliable
form for rare gases turns out to be the (modified) Buckingham or exp-6
potential~\cite{Buckingham} (see it defined at Eq.\,(\ref{2-1})), as
parametrized through high-density experimental data~\cite{Ross}.
As a matter of fact, when HT/HP conditions hold,
an exponential law is a more adequate representation for
the interatomic repulsion at short distances than is a power law.
In this respect, we can say that the exp-6 potential takes into account
the effects of the three-body interaction in an isotropic fashion.

Recently, some controversy has arisen about the topology of the HT/HP
part of the phase diagram of Xe, whose behaviour should be representative
also of Ar and Kr.
Stimulated by the findings of a recent laser-heated diamond-anvil-cell
(DAC) experiment on Xe~\cite{Boehler}, Belonoshko {\it et al.} have done
molecular-dynamics (MD) computer simulations of the exp-6 potential in
order to interpret those data~\cite{Belonoshko1,Belonoshko2}.
In the DAC experiment, a Xe fluid sample is compressed at a given pressure
in the range between 10 and 40 GPa and subsequently heated with a laser
until a bulk transition is detected.
The experiment is meant to provide information about the freezing of
fluid Xe in the HT/HP region and, ultimately, to draw the fluid-solid
coexistence line in the $P$-$T$ diagram.
Surprisingly, the freezing line shows an unexpected cusp near
$T=2700$\,K and $P=15$\,GPa, which was originally imputed to the
boundary between a fluid-solid and a fluid-glass transition~\cite{Boehler}.
The numerical simulations of the exp-6 model by Belonoshko {\it et al.}
have instead revealed a competition between two distinct, FCC and BCC solid
phases, which happen to exchange their relative thermodynamic stability
right in the region of the observed cusp.
More precisely, for temperatures above $T\simeq 2700$\,K, a fluid-BCC-FCC
sequence of phases is reported upon isothermically increasing the pressure
beyond 25 GPa.
This interpretation has been criticized by Kechin~\cite{Kechin}.

The simulation method used in Refs.\,\cite{Belonoshko1,Belonoshko2} was the
so-called two-phase or coexistence method~\cite{Morris1,Morris2},
as implemented into an isothermal-isobaric MD simulation~\cite{Belonoshko3}.
In point of principle, this technique should be able to detect a coexistence
between two structurally distinct phases whenever it occurs; in practice,
this method has some limitations since it requires large sizes and very
long simulation times to reach good equilibration in the region of the
interface between the coexisting phases~\cite{Agrawal}.
Moreover, it cannot be excluded that the route to equilibrium of an
unstable interface will pass through (and long remain in) a phase which,
though not being the preferred one under bulk conditions, is nonetheless
promoted by (small) spatial inhomogeneities of the system.
This would be especially so when the difference in Gibbs free energy between
two distinct solid phases is very small: in that case, a system being
prepared in a metastable solid phase would take very long time (possibly
much longer than accessible to simulation) to relax to the truly stable solid.

In order to gain further insight into the nature of Xe freezing at the
temperatures investigated by the DAC experiment, we have carried out a
Monte Carlo simulation in the canonical ensemble of the same exp-6 system
of Refs.\,\cite{Belonoshko1,Belonoshko2}, but now calculating the
free energy of the relevant solid phases by the method of Frenkel
and Ladd~\cite{Frenkel1}, which would provide the benchmark for
solid-free-energy evaluations.
When used in conjunction with conventional thermodynamic integration,
this method would allow one to draw the ``exact'' phase diagram of the
given model system and the only source of error would be associated with
the finite size of the sample.
Indeed, if the chemical-potential gaps between different solid structures
are minute, also the statistical errors affecting the relevant thermodynamic
averages are to be made sufficiently small ({\it i.e.}, by carrying out
long enough simulation runs).
As a matter of fact, we do neither confirm the findings of Belonoshko
{\it et al.} -- for we predict that the BCC solid would become stable
in Xe only above 4500 K -- nor find any good agreement with the DAC data
of Ref.\,\cite{Boehler} at the highest temperatures, {\it i.e.}, no clear
bump is seen along the freezing line.
A similar outcome was also found by Frenkel for He~\cite{Frenkel2},
whose BCC-FCC transition would only occur in a region of the phase
diagram where the fluid is stable.

The outline of the paper is the following: in Section 2, we introduce
the model system and describe our simulation method.
In Section 3, we present our numerical results and draw the ensuing
phase diagram. Then, in Section 4, we make a critical comparison of
our outcome with those of previous studies.
Further comments and remarks are postponed to the Conclusions.

%
%
\section{Model and method}
\renewcommand{\theequation}{2.\arabic{equation}}
\subsection{The exp-6 model}
 
All textbooks in statistical mechanics report that rare-gas thermal
behaviour is well accounted for by the simple Lennard-Jones pair potential.
However, it is less known that, when rare gases are {\em very dense}, a
different potential with a softer repulsive shoulder is better suited.
Such is the exp-6 potential, defined to be
\be
v(r)=\left\{ 
\begin{array}{ll}
+\infty\,, & r<\sigma \\
\frac{\epsilon}{\alpha-6}
\left\{6\exp\left[\alpha\left(1-\frac{r}{r_m}\right)\right]-
\alpha\left(\frac{r_m}{r}\right)^6\right\}\,, & r\geq\sigma
\end{array}
\right. \,,
\label{2-1}
\ee
where $\alpha$ controls the softness of the repulsion and $\epsilon>0$
is the depth of the potential minimum located at $r_m$.
We select the value of $\sigma$ in such a way that, for the given
$\alpha$ and $r_m$, the function appearing in the second line of
Eq.\,(\ref{2-1}) reaches its maximum right at $\sigma$.
Hence, as $r$ moves down to $\sigma$, $v(r)$ reaches a stationary
value and then goes abruptly to infinity.
Ross and Mc Mahan have shown that, for a suitable choice of the
parameters $\alpha,\epsilon$, and $r_m$, $v(r)$ gives a rather
faithful representation of many thermodynamic properties of Ar
at high densities.
Moreover, a corresponding-state theory appears to hold, which permits
to derive the model parameters appropriate to Kr and Xe from those of Ar.
For Xe, such best parameters are $\alpha=13$, $\epsilon/k_B=235$\,K,
and $r_m=4.47$\,\AA~\cite{Ross}, where $k_B$ is the Boltzmann constant
(for $\alpha=13$, the value of $\sigma$ is then $0.246972\ldots r_m$).
In the following, we work with dimensionless variables, {\it i.e.},
$T^*=k_BT/\epsilon$, $P^*=Pr_m^3/\epsilon$, and $\rho^*=\rho r_m^3$
($\rho\equiv N/V$ is the particle-number density).
The accuracy of the exp-6 modelization for Xenon at low pressures was
tested by Belonoshko {\it et al.} against available experimental
results and, up to 0.7 GPa, was found to be good (see Fig.\,7 of
Ref.\,~\cite{Belonoshko1}).

Recently, the vapour-liquid equilibria of the exp-6 fluid were
studied numerically through a series of Monte Carlo (MC) simulations
in the Gibbs ensemble~\cite{Tavares}.
As a result, the critical point was located at $\rho^*=0.303$ and
$T^*=1.316$.
From a simulation study by Errington and Panagiotopoulos a slightly
different estimate of the critical-point coordinates is extracted,
namely $\rho^*=0.318$ and $T^*=1.318$~\cite{Errington}.
Finally, the freezing density of the exp-6 fluid was estimated for a
number of temperatures by the Hansen-Verlet criterion~\cite{Vortler,Lizal}.

\subsection{Details of the Monte Carlo simulation}

The numerical-simulation method provides virtually exact information on
the statistical behaviour of a given model system.
We have performed Monte Carlo (MC) simulations of the exp-6 model in
the canonical ensemble ({\it i.e.}, at constant temperature $T$,
volume $V$, and number $N$ of particles), using the standard
Metropolis algorithm for sampling the equilibrium distribution in
configurational space.
The values of $N$ that we only consider are those that fit to
a cubic simulation box with an integer number of cells, {\it i.e.},
$N=4n^3$ for the FCC solid and $N=2n^3$ for the BCC solid, $n$ being
the number of cells along any spatial direction.
For a given particle number, the length $L$ of the box is adjusted
to a chosen density value $\rho$, {\it i.e.}, $L=(N/\rho)^{1/3}$.
Called $a$ the distance between two nearest-neighbour reference
lattice sites, we have $a=(\sqrt{2}/2)(L/n)$ for a FCC crystal
and $a=(\sqrt{3}/2)(L/n)$ for a BCC crystal.
Finally, periodic conditions are applied to the box boundaries.

Our largest sample sizes were $N=1372$ (FCC) and $N=1458$ (BCC) which,
to all practical purposes, can be regarded as nearly the same size.
We point out that comparing the statistical properties of similar (large)
sizes is mandatory when we want to decide which solid phase is stable at
given $T$ and $P$, since this produces comparable statistical errors for
the two phases.
Otherwise, one can resort to some extrapolation to the thermodynamic limit
which is however computationally far more demanding.

Some X-ray diffraction studies indicate that the FCC-HCP transition
could be a common behaviour in all rare-gas
solids~\cite{McMahan,Jephcoat2,Errandonea}.
At room temperature, the HCP solid would be stable above 75 GPa, while
metallization is not expected before $\approx 130$ GPa~\cite{Eremets}.
We have thus carried out also a number of simulations of a HCP crystal
hosting $10\times 12\times 12=1440$ particles (the simulation box
is only approximately cubic in this case and the $z$ axis is now
oriented along the $[111]$ direction
-- the three box sides have lengths of $L_x=n_xa$,
$L_y=(\sqrt{3}/2)n_ya$, and $L_z=(\sqrt{6}/3)n_za$,
with $n_x=10$, $n_y=n_z=12$, and $a=(\sqrt{2}/\rho)^{1/3}$).

To locate the transition point at a given $T$, we have numerically
generated two isothermal quasistatic paths, starting from the very
dilute fluid on one side and from the highly compressed solid on the
other side.
As a rule, in the solid region, the last MC configuration produced at
a given $\rho$ serves, after suitable rescaling of particle coordinates,
as the starting configuration for the run at a slightly lower density.
Similarly, in the dense-fluid region, the simulations are carried out
in a chain, {\it i.e.}, the run at a given density is started from the
last (rescaled) configuration produced at a lower $\rho$ value.
The FCC and BCC solid paths are followed until the fluid spontaneously
forms during the MC run, as evidenced by the abrupt change in energy
and pressure.
Usually, we were able to overheat the solid for a little beyond the
fluid-solid phase boundary, while undercooling of the fluid is much
easier.
For each $\rho$ and $T$, equilibration of the sample typically
takes $2\times 10^3$ MC sweeps, a sweep consisting of one attempt
to sequentially change the position of all particles.
The maximum random displacement of a particle in a trial MC move is
adjusted once a sweep during the run so as to keep the acceptance
ratio of the moves as close to 50\% as possible, with only small
excursions around this value.

For given $NVT$ conditions, the relevant thermodynamic averages are
computed over a trajectory whose length ranges from $2\times 10^4$
to $6\times 10^4$ sweeps.
The excess energy per particle $u_{\rm ex}$, the pressure
$P$, and (in the solid phase) the mean square deviation $\delta R^2$ of a
particle from its reference lattice position are especially monitored.
Pressure comes from the virial formula,
\be
P=\rho k_BT+\frac{\left<{\cal V}\right>}{V}\,,\,\,\,
{\cal V}=-\frac{1}{3}\sum_{i<j}r_{ij}v'(r_{ij})
\label{2-2}
\ee
($r_{ij}$ is the distance between particles $i$ and $j$).
In practice, to avoid double counting of interactions, the pair
potential is truncated above a certain cutoff distance $r_c$, which
is taken to be only slightly smaller than $L/2$.
Then, the appropriate long-range corrections are applied to energy
and pressure by assuming $g(r)=1$ beyond $r_c$, $g(r)$ being the
radial distribution function (RDF).

The RDF histogram is constructed with a spatial resolution of
$\Delta r=r_m/50$ and updated every 10 MC sweeps.
The RDF is computed up to a distance of $L/2$: at that distance,
the $g(r)$ was never found to significantly differ from unity, at
least for the largest system sizes.

To evaluate the numerical errors affecting the main statistical
averages, we divide the MC trajectory into ten blocks and estimate
the length of the error bars as being twice the empirical standard
deviation of the block averages from the mean (under the implicit
assumption that the decorrelation time of any relevant variable is
less that the size of a block).
Tipically, the relative errors of energy and pressure are of few tenths
of a percent.

The difference in excess free energy between two equilibrium states
of the system, say 1 and 2, lying {\em within the same phase} is
computed through the combined use of the formulae
\be
\frac{f_{\rm ex}(T_2,\rho)}{T_2}=\frac{f_{\rm ex}(T_1,\rho)}{T_1}-
\int_{T_1}^{T_2}{\rm d}T\,\frac{u_{\rm ex}(T,\rho)}{T^2}
\label{2-3}
\ee
and
\be
\beta f_{\rm ex}(T,\rho_2)=\beta f_{\rm ex}(T,\rho_1)+
\int_{\rho_1}^{\rho_2}{\rm d}\rho\,\frac{1}{\rho}
\left[\frac{\beta P(T,\rho)}{\rho}-1\right]\,,
\label{2-4}
\ee
where $f_{\rm ex}$ is the excess Helmholtz free energy per particle and
$\beta=(k_BT)^{-1}$.
The integrals in Eqs.\,(\ref{2-3}) and (\ref{2-4}) are performed
numerically by applying the Simpson rule to a linear-spline
approximant of the simulation data for energy and pressure.

The above formulae are however useless if one does not have an
independent estimate of the system free energy in a reference state.
Only in this case, Eqs.\,(\ref{2-3}) and (\ref{2-4}) help finding
the free energy of any other state in the same phase.
The choice of such a reference state is different for the fluid and
for the solid phase.
As a reference state for the fluid, we can choose any equilibrium state
been characterized by a very small $\rho$ value and arbitrary $T$ (say, a
nearly ideal gas), since then the excess chemical potential of the system
can be accurately estimated by the Widom or particle-insertion method,
\be
\mu_{\rm ex}=-k_BT\ln\left<\exp\left(-\beta E_{\rm ins}\right)\right>\,,
\label{2-5}
\ee
where $E_{\rm ins}$ comprises all interaction terms between a
randomly-inserted ghost particle and all the system particles.
The average in Eq.\,(\ref{2-5}) is evaluated numerically during a
run of typically $5\times 10^4$ equilibrium sweeps, with an insertion
attempted at the completion of every sweep.
Once $\mu_{\rm ex}$ is given, the excess values of free energy and
entropy will follow from
\be
\beta f_{\rm ex}=\beta\mu_{\rm ex}-\frac{\beta P}{\rho}+1
\,\,\,\,\,\,{\rm and}\,\,\,\,\,\,
\frac{s_{\rm ex}}{k_B}=\beta(u_{\rm ex}-f_{\rm ex})\,.
\label{2-6}
\ee
It is useful to note that, from a strictly numerical point of view,
choosing a very dilute gas as a reference state for the fluid is
far better than starting the thermodynamic integration in
Eq.\,(\ref{2-4}) from the ideal gas of equal temperature.
In fact, unless one has a lot of thermodynamic points in the very-dilute
region of the phase diagram, a spline interpolant of $\beta P/\rho$ that
is sufficiently accurate in this region is hard to construct.

\subsection{The Frenkel-Ladd calculation of solid free energies}

The Frenkel-Ladd (FL) method for calculating the free energy of a
solid system relies on a different kind of thermodynamic integration%
~\cite{Frenkel1,Polson}.
The idea is to continuously transform the system of interest with
potential energy $V_1$ into a reference solid system of known free
energy (typically, an Einstein crystal).
This is accomplished through a linear interpolation of the
potential-energy functions $V_0$ and $V_1$ of the two systems, {\it i.e.},
$V_\lambda=V_0+\lambda(V_1-V_0)$ (with $0\le\lambda\le 1$).
Then, the difference in free energy between two homologous $NVT$ states
of the systems is calculated via the exact Kirkwood formula
\be
F_1-F_0=\int_0^1{\rm d}\lambda\left<V_1-V_0\right>_\lambda\,,
\label{2-7}
\ee
where the average $\left<\ldots\right>_\lambda$ is taken over the
equilibrium distribution of the hybrid system with $V_\lambda$
potential and is evaluated numerically by a Monte Carlo procedure.

Upon denoting $\{{\bf R}_i^{\rm (0)},i=1,\ldots,N\}$ the reference lattice
positions, the Einstein model is described by an interaction potential of
\be
V_{\rm Ein}({\bf R}^N)=\frac{1}{2}c\sum_{i=1}^N
\left({\bf R}_i-{\bf R}_i^{\rm (0)}\right)^2\,,
\label{2-8}
\ee
where $c$ is the spring constant -- the same for all oscillators.
It readily follows from Eq.\,(\ref{2-8}) that the free energy and mean
square separation of a particle from its reference lattice site are given by
\be
\beta F_{\rm Ein}=-\frac{3N}{2}\ln\left(\frac{2\pi}{\beta c\Lambda^2}\right)
\,\,\,\,\,\,{\rm and}\,\,\,\,\,\,
\delta R^2_{\rm Ein}=\frac{3}{\beta c}\,,
\label{2-9}
\ee
where $\Lambda$ is the thermal wavelength.
In deriving the first of Eqs.\,(\ref{2-9}), no Gibbs factor of $N!$ was
included in the partition function since the Einstein oscillators are
distinguishable entities.

A non trivial problem with this choice of reference system, already
pointed out in the original article \cite{Frenkel1}, is the following:
while the Einstein particles can only perform limited excursions around
their reference positions, which implies a finite value of $\delta R^2$,
there is no means to constrain particles interacting through $V_1$ to
remain confined in the neighborhood of their initial positions, even in
the solid phase.
In other words, at variance with $V_{\rm Ein}$, $V_1$ is a
translationally-invariant potential, with the result that the integrand
in Eq.\,(\ref{2-7}) shows a divergence, in the thermodynamic limit, at
$\lambda=1$.
To overcome this problem, Monte Carlo simulations of $V_\lambda$ are
usually performed under the constraint of a fixed center of mass.
We do not repeat here the entire analysis showing how to deal with such
a constraint in the equilibrium sampling, but simply quote the final
result (see Ref.\,\cite{Polson} for details):
\ba
\beta f_{\rm ex} &\equiv& \beta\frac{F-F_{\rm id}}{N}
\nonumber \\
&=& -\ln\left(\rho(\beta c)^{-3/2}\right)-\frac{3}{2}\ln(2\pi)+1-
\frac{2\ln N}{N}+\frac{\ln(2\pi)}{N}+
\frac{\ln\left(\rho(\beta c)^{-3/2}\right)}{N}
\nonumber \\
&+& \frac{\beta}{N}\int_0^1{\rm d}\lambda
\left<\Delta V\right>_\lambda^{\rm CM}\,,
\label{2-10}
\ea
where the superscript CM denotes a constrained average and $\Delta V=V_1-V_0$.
Polson {\it et al.} have conjectured that, on fairly general grounds,
$\beta f_{\rm ex}(N)+\ln N/N=
\beta f_{\rm ex}(\infty)+{\cal O}(N^{-1})$~\cite{Polson}.
We shall see later whether this is found in our case.
Given the value of $f_{\rm ex}$, the excess chemical potential of the
solid still follows from the first of Eqs.\,(\ref{2-6}).

A few considerations on the numerical implementation of the FL method
are now in order.
While in principle the value of $c$ can be chosen arbitrarily, just for
numerical purposes it is convenient to take it such that $\delta R^2$ of
the target solid is close to $3/(\beta c)$: in this case, the variations
of the integrand in Eq.\,(\ref{2-7}) with $\lambda$ are likely to be small
as well as the error in performing the numerical quadrature.

A further caveat must be added for hard-core potentials (as is the one
we are interested in here).
If we write the potential as the sum of a hard-sphere part plus a remainder,
$V_1=V_{\rm HS}+V_r$, then it is mandatory to include a $V_{\rm HS}$ term
also in $V_0$, since otherwise there is no way to transform continuously
from $V_1$ to $V_0$.
As a result,
\be
V_\lambda=V_{\rm HS}+V_{\rm Ein}+\lambda(V_r-V_{\rm Ein})\,,
\label{2-11}
\ee
implying $\Delta V=V_r-V_{\rm Ein}$ in Eq.\,(\ref{2-10}).
However, the free energy of a system of hard-core (that is, mutually
interacting) Einstein particles is not known; what we can only say is
that this interaction cannot have any consequence on the thermodynamic
quantities of an Einstein crystal in so far as $c$ takes sufficiently
large values, since in this case particles are blind to each other.
Then, unless $c$ is given a very huge value, the FL calculation
is only viable for a very cold solid.

Whether a value of $c$ is huge or not can be decided from the comparison
between $\sqrt{3/(\beta c)}$ and $(a-\sigma)/2$.
Should the former be much smaller than the latter, the error committed
in assuming the form (\ref{2-9}) for the free energy of the interacting
Einstein crystal is presumably negligible.
Frenkel and Ladd have indicated a more rigorous method to ascertain the
importance of hard-core interactions for the free energy of an Einstein
solid, based on the estimate of the leading (virial) correction to the
free energy of the non-interacting Einstein particles~\cite{Frenkel1}.

%
%
\section{Results}
\renewcommand{\theequation}{3.\arabic{equation}}
\setcounter{equation}{0}

\subsection{Test calculations}

We have first checked our FL simulation code against the
hard-sphere calculation given in Ref.\,\cite{Polson}.
We take a system of $N=256$ hard spheres of diameter $\sigma$
at $\rho\sigma^3=1.0409$ (this is about the melting density of the
FCC solid), with $c=6000\,k_BT/\sigma^2$.
We estimate numerically the average of $\Delta V=-V_{\rm Ein}$
over the canonical distribution relative to $V_\lambda$ for a number
of $\lambda$ values in the range from 0 to 1.
Precisely, we take $\lambda$ to increase in steps of 0.05 from 0 to 0.9,
with a smaller step of 0.01 in the range from 0.9 to 1.
For each value of $\lambda$, as many as $2\times 10^4$ MC sweeps are
produced at equilibrium.
The MC moves are such that the position of the system center of mass
is fixed; to the same purpose, if a particle happens to move (slightly)
out of the simulation box, no attempt is made to put it back into the
box~\cite{Frenkel3}.
Besides the value of $\left<\Delta V\right>_\lambda$, we also estimate
the statistical error affecting this quantity by partitioning the MC
trajectory into large blocks.
Finally, the integral over $\lambda$ in Eq.\,(\ref{2-10}) is
calculated using the Simpson rule.
For the chosen $c$ value, the virial correction to the free energy
of the Einstein crystal due to hard-core interaction between particles
is negligible ($\beta\Delta f_{\rm ex}\approx 10^{-8}$).
We thus find $\beta f_{\rm ex}+\ln N/N=5.891(5)$ for the FCC solid,
which is fully consistent with the result obtained by Polson
{\it et al.} (see Fig.\,2 of Ref.\,\cite{Polson}).

As a further check of our numerical code, we have considered a model
system of soft spheres interacting through the repulsive potential
$v(r)=\epsilon\left(\sigma/r\right)^{12}$.
We take a system of $N=256$ particles with $k_BT/\epsilon=1$,
$\rho\sigma^3=1.1964$, and $c=500\,\epsilon/\sigma^2$.
We estimate numerically the average of
$\Delta V=V_1-(V_1^{(0)}+V_{\rm Ein})$ over the canonical distribution
relative to $V_\lambda$ for a number of $\lambda$ values in the range
from 0 to 1 ($V_1^{(0)}$ is the total potential energy as calculated
for particles located in the respective perfect-lattice positions).
For such $\Delta V$, the quantity $\beta V_1^{(0)}/N$ must be added
to the r.h.s. of Eq.\,(\ref{2-10}) to obtain $\beta f_{\rm ex}$.
For each value of $\lambda$, as many as $5\times 10^4$ MC sweeps
are produced at equilibrium.
In the end, we find $\beta f_{\rm ex}+\ln N/N=9.208(2)$ for the FCC solid,
which again perfectly agrees with Fig.\,1 of Ref.\,\cite{Polson}.

\subsection{The exp-6 simulation}

Moving to the exp-6 model, we have calculated the excess free energy
of both the FCC and the BCC solid for a number of $(\rho^*,T^*)$ pairs,
using for the integral in Eq.\,(\ref{2-10}) the same specifications as
given above for the test calculation on hard spheres.
Though one single free-energy calculation by the FL method would
be sufficient for estimating the free energy of the solid in any
other state, we have often found more practical to repeat the FL
calculation rather than generating a long thermodynamic path to a
very distant point in the phase diagram.
In Table 1, we collect the calculated values of the excess free energy
per particle in a few state points, for three different solid phases
of the exp-6 model (and for just the largest sizes).
The numerical precision reported for each $\beta f_{\rm ex}$ is the
Simpson integral of the errors that are associated with the values
of $\beta\left<\Delta V\right>_\lambda^{\rm CM}/N$, as calculated
for the selected $\lambda$ points~\cite{note}.
In the same table, we indicate the value of $c^*=cr_m^2/\epsilon$ that
is considered for each single state.
In every case, this $c$ is such that $3/(\beta c)\simeq\delta R^2$
of the exp-6 solid.
Since all of the tabulated state points lie sufficiently far from the
melting line, the average square excursion of an exp-6 particle from
its reference lattice site turns out to be quite small.
Hence, all of the $c^*$'s are much larger than 1 and the virial correction
to the Einstein-crystal free energy is absolutely negligible.

We get an indirect check of the free-energy values in Table 1 by
calculating, via ordinary thermodynamic integration based on
Eqs.\,(\ref{2-3}) and (\ref{2-4}), the difference in excess free energy
between various pairs of BCC and FCC states in the table.
We have always found a perfect agreement with the tabulated values,
to within the reported numerical precision, for both types of solids.
For each $(\rho^*,T^*)$ pair in Table 1, we have finally verified that
the linear scaling of $\beta f_{\rm ex}(N)+\ln N/N$ as a function of
$N^{-1}$ holds well, as demonstrated in Fig.\,1 in one case only.

In order to see whether a stable BCC phase exists above $T^*\simeq 11$
(which is where the authors of Ref.\,\cite{Belonoshko1} would locate
the fluid-BCC-FCC triple point), we have carried out extensive MC
simulations of the exp-6 model for a number of $T^*$ values
($4.25,8.15,12.77,16,20$, and 25).
For all such values, we construct the fluid path and two solid, FCC and
BCC paths (for $T^*=4.25$, only the FCC equation of state is generated).
For $T^*=16,20$, and 25, we have also constructed the equation of state for
$N=1440$ particles in a HCP arrangement.
With the only exception of $T^*=4.25$, where $N=864$, the other isotherms
are plotted for $N=1372$ (fluid and FCC) and $N=1458$ (BCC) particles.
Once the free energy of the system is known along an isothermal path,
the chemical potential $\mu$ along the same path readily follows from
the first of Eqs.\,(\ref{2-6}) as either a function of $\rho$ or,
equivalently, of $P$.
For given $T$ and $P$ values, the thermodynamically stable phase is the one
with lower $\mu$: hence, the phase transition from fluid to {\it e.g.}
FCC at constant temperature is located at the pressure where the chemical
potential of the fluid takes over the $\mu$ of the FCC solid.

Surprisingly, we find that, at variance with the conclusions of
Ref.\,\cite{Belonoshko1,Belonoshko2}, the BCC phase of the exp-6
model is only stable at very high temperatures, {\it i.e.}, above
$T^*\simeq 20$, and in a narrow pressure interval.
For all reduced temperatures up to 16, the scenario is similar
(see it represented in Figs.\,2 and 3 for $T^*=16$): as pressure
grows, the chemical potential of the fluid eventually overcomes
that of the FCC solid; at the crossing point $P_{\rm FS}$, the
latter is already lower than the BCC chemical potential.
The BCC solid would be more stable than the FCC solid only for
pressures lower than $P_{\rm SS}$ ($<P_{\rm FS}$), that is in
a region where the fluid is the stable phase
(however, at $T^*=4.25$, a pressure $P_{\rm SS}$ simply does not exist,
{\it i.e.}, the BCC solid is never preferred to the FCC solid).
Values of $P_{\rm SS},P_{\rm FS}$, and of
$\Delta\mu\equiv\mu_{\rm FCC}-\mu_{\rm BCC}$ (as calculated at
the fluid-solid transition pressure $P_{\rm FS}$) are reported
in Table 2 for the various temperatures.
In the same table, the values of the freezing and the melting
densities are also indicated.
All the tabulated quantities are reported with three decimal digits,
with no indication of the estimated statistical errors.
The general question of the numerical reliability of our results is
postponed to the next Section, but we can anticipate that the results
are fully trustworthy.

Fig.\,2 shows the chemical potential excess $\Delta\mu(P)$ of the FCC
phase relatively to BCC, for $T^*=16$.
This quantity is larger than zero ({\it i.e.}, the stable solid would be
of BCC structure) only inside the fluid region of the phase diagram.
Hence, the fluid directly transforms into the FCC solid, while the BCC
phase is metastable.
In the same picture, we also plot the difference in chemical potential
between FCC and HCP, a quantity which is nearly constant and close to
$-0.004$ at all pressures.
This means that the HCP solid is never stable, albeit it is about to be
so at all pressures.

Considering the small $\mu$ gaps between the various solids, one may
wonder whether the above conclusions are influenced by the finite
size of the simulated samples.
To this purpose, we have taken $T^*=16$ and plotted in Fig.\,2 the same
quantity $\Delta\mu(P)$ as before, but now calculated for a smaller FCC
solid of 864 particles and for a BCC solid of 1024 particles.
The new curve differs from the old one just for a small rigid shift
towards low pressures, and crosses zero at
$\beta P_{\rm SS}=67.686\,r_m^{-3}$ (to be contrasted with the previous
value of $69.130\,r_m^{-3}$).
A bit smaller is the variation of $\beta P_{\rm FS}$ as $N$ changes from
1372 to 864 (it moves from $71.877\,r_m^{-3}$ to $71.051\,r_m^{-3}$).
We expect that similar results would be obtained at the other
temperatures.
Considering the large $N$ values involved, it is very unlikely that our
conclusions for the largest sizes can be reversed in the thermodynamic limit.

In Fig.\,3, the various branches of the equation of state $\beta P$
vs. $\rho$ are plotted for $T^*=16$.
This curve shows a straight-line cut at $P=P_{\rm FS}$, which allows
one to identify the values of $\rho_{\rm f}$ and $\rho_{\rm m}$ for
the given $T$.
Correspondingly, the Helmholtz free energy shows, as a function of the
specific volume $v=1/\rho$, a straight-line behaviour in the interval
between $\rho_{\rm m}^{-1}$ and $\rho_{\rm f}^{-1}$.
This line is the common tangent of the fluid and the FCC free-energy
curves (its equation is $\beta f=\beta\mu_{\rm FS}-\beta P_{\rm FS}v$).
Finally, a look at the RDFs in the coexistence region indicates that
the typical values of the nearest-neighbour distance in all phases
fall well within the repulsive shoulder of the pair potential,
as expected for any highly-compressed system.

An inspection of Table 2 shows that, as temperature grows, the
difference $P_{\rm FS}-P_{\rm SS}$ gradually reduces until it would
cross zero at $T^*\approx 19$, as being suggested by a power-law
extrapolation beyond $T^*=16$ of $P_{\rm SS}(T)$ and $P_{\rm FS}(T)$.
Originally, it was just this evidence that led us to include in our
calculations two other isotherms at $T^*=20$ and 25, in order to see
whether a BCC-FCC phase transition eventually shows up at very high
temperatures.
Indeed, when $T^*$ is as high as 20, we find that, on increasing
pressure, the sequence of stable phases of the exp-6 model is
fluid-BCC-FCC, with transitions at $\beta P_{\rm FS}=83.517\,r_m^{-3}$
(fluid-to-BCC transition) and at $\beta P_{\rm SS}=83.575\,r_m^{-3}$
(BCC-to-FCC transition).
Actually, at this temperature the BCC window is so narrow that the
fluid-BCC-FCC triple temperature should be very close to $T^*=20$.
At $T^*=25$, the interval of stability of the BCC phase is much wider
($\approx 6$ GPa).
Table 3 reports thermodynamic data for $T^*=20$ and 25, whereas the
difference in chemical potential between the various phases is
shown in Fig.\,4 for $T^*=25$ as a function of pressure.
Hence, the BCC phase becomes eventually stable, but at much higher
temperatures than estimated in Ref.\,\cite{Belonoshko1}.

In the end, the $P$-$T$ phase diagram of the exp-6 model would appear
as in Fig.\,5, where temperature and pressure are reported in Xe units.
In this picture, our fluid-solid coexistence data are contrasted with
the coexistence loci of Belonoshko {\it et al.} and with the Simon-Glatzel
equation of state for Xe, as deduced from that of Ne by the simple
rescaling of pressure and temperature values proposed by Vos
{\it et al.}~\cite{Vos}.
Compared to the findings of Ref.\,\cite{Belonoshko1}, freezing of the
exp-6 system occurs for slightly lower pressures in our simulation; more
important, our fluid-FCC-BCC triple point lies very far from the location
indicated by Belonoshko {\it et al.}

To better understand the nature of the differences between our results
and those of Belonoshko {\it et al.}, we have plotted in Fig.\,6 the
exp-6 phase diagram on the $T$-$V$ plane.
It can be appreciated from this picture that small differences in the
volumes of the coexisting fluid and FCC solid are found at all
temperatures.
On the contrary, isobaric paths are pretty much the same (see Fig.\,11
of Ref.\,\cite{Belonoshko1} for a comparison), indicating that where we
deviate from Belonoshko {\it et al.} is essentially in the definition
of a phase-coexistence condition, not in the calculation of basic
statistical averages.

Where we radically contrast with Belonoshko {\it et al.} is really in
the position of the BCC-FCC transition line, with ours running much
higher in temperature and pressure.
To assess how much distant is our prediction from these authors,
we note that, for {\it e.g.} $T^*=16$, they report a reduced pressure
of about 70 GPa ($P^*/T^*\simeq 120$) at the BCC-FCC transition while,
at the same pressure, our $\beta\Delta\mu$ is about $-0.040$ (see
Fig.\,2), {\it i.e.}, appreciably larger, in absolute terms, than
the size of statistical errors.

%
%
\section{Discussion}
\renewcommand{\theequation}{4.\arabic{equation}}
\setcounter{equation}{0}

Having devoted the last Section to a plain presentation of our results,
we now reconsider more critically their implications, especially in
relation to the numerical precision of the simulation data.

The crucial quantity to look at is, obviously, the chemical potential.
One source of error in its calculation follows from the finite size of
the simulated system, a problem further complicated by the impossibility
to compare FCC and BCC solids with equal numbers of particles.
In point of principle, this would force us to some sort of extrapolation
to $N=\infty$ which, however, is computationally demanding.
Instead, we have decided to simulate fluid, FCC, BCC, and HCP samples of
similar size, which are also sufficiently large that no significant
finite-size error would be made in tracing the coexistence lines.
Indeed, we have already demonstrated in the previous Section -- see the
comment to Fig.\,2 -- that this kind of error is not very important.

Further errors in estimating the relative stability of two distinct
solid phases are imputable to the limited precision with which we
compute the excess free energy of the reference solid states in
Table 1 as well as the pressure values along an isothermal path.
Looking at Table 1, the overall error on the free-energy difference
between BCC and FCC in units of $k_BT$ is of about three units on the
third decimal place, and comparable is the statistical error
accompanying the values of $\beta P/\rho$ for each phase.
The error associated with the $\beta\mu$ gap between the fluid and a
solid phase is again of $4\div 5\times 10^{-3}$, but the rate at which
this quantity varies as a function of pressure is much larger
(this implies that the location of the fluid-solid transition is
much better defined numerically than is the solid-solid transition point).
Summing up, we expect that the typical statistical error affecting
the BCC-FCC $\beta\Delta\mu$ is of about $10^{-2}$.
If this is true, it is evident from the tabulated values of
$\beta\Delta\mu$ that we cannot definitely rule out the possibility
that the BCC solid becomes stable at temperatures lower than $T^*=20$,
and even as small as 11.
Only if we turn to much longer Monte Carlo runs than hereby considered,
we can hope to reduce drastically the width of the error bars.
Anyway, even in the worst case, we can safely infer from our data that
the BCC phase may only be stable in a narrow slice of few GPa adjacent
to the fluid-BCC coexistence locus, a pressure interval much narrower
than predicted by Belonoshko {\it et al.}

The caution expressed above is probably overstated: it is a fact that
all the curves plotted in Figs.\,2 and 4 are very smooth, which suggests
that the statistical noise underlying the profile of $\beta\Delta\mu$ is
quite smaller than the {\em maximum} estimated above.
In this case, the conclusions drawn in the previous Section just on the
basis of the {\em average} behaviour of $\Delta\mu$ are substantially
correct, and the BCC phase would really be metastable below $T^*\approx 20$.
If we believe this, a stable BCC phase would first appear in Xe at so high
a temperature ($\approx 4500$\,K) that one can even wonder whether the
BCC-FCC transition in Xe is preempted by quantum effects.

As a matter of fact, our simulation results do not match the experimental
data of Boehler {\it et al.} at the highest temperatures.
Even assuming that the exp-6 system is a very good representation of Xe
in the HT/HP regime, the problem could be in fact with the experiment,
which might be far from realizing hydrostatic conditions.
In fact, it is generally believed that, in a laser-heated DAC experiment,
the stress state within the diamond cell may not be hydrostatic; moreover,
the sample can even exhibit considerable shear stresses~\cite{Kavner}.
In consideration of the small difference in chemical potential between
BCC, FCC, and HCP, it could then be that what is experimentally recognized
as solid is in fact a mixture of BCC and FCC/HCP crystallites.

%
%
\section{Conclusions}

In this paper, we have reported on the results of an extensive $NVT$
Monte Carlo simulation of the exp-6 model, an effective-interaction
model that is thought to provide a realistic description of the
thermal properties of rare gases under extreme,
high-temperature/high-pressure conditions.
We have plotted the ``exact'' phase diagram of this system upon
combining the method of thermodynamic integration with fully-fledged
free-energy calculations both in the fluid (by the Widom method) and
in the solid phase (by the Frenkel-Ladd method).

The aim of this effort was to point out some aspects concerning the
uncertain status of the solid phase of Xenon at high densities.
In a previous molecular-dynamics study of the exp-6 system, with
parameters being appropriate to Xe, the two-phase method was employed
for simulating the thermodynamic coexistence between fluid and solid%
~\cite{Belonoshko1,Belonoshko2}.
This study gave evidence of a stable BCC phase in a narrow range of
pressures above 25 GPa, when temperature exceeds $2700\,$K, {\it i.e.},
near to where a laser-heated, diamond-anvil-cell experiment finds
a cusp on the freezing line of Xe.
In fact, our free-energy calculations partially contradict such
findings, showing that Xe freezes directly into a FCC solid, the
BCC phase becoming stable only at much higher temperatures
(above $4500\,$K), provided that the exp-6 modelization is still
valid for Xe in these extreme conditions;
moreover, the HCP solid is practically as stable
as the FCC one.

A clue to understand this disagreement is the very small difference in
Gibbs free energy, for $T>2500\,$K, between the BCC and FCC solids of
equal pressure and similar size.
The little advantage of FCC over BCC at freezing could be the reason
for the apparent stability of the BCC phase in the former numerical
study of the exp-6 model.
In our simulation study, we find that, at sufficiently high temperature,
there is a narrow range of pressures where the BCC solid is more stable
than the FCC, but less stable than the fluid.
Only beginning from $T^*=20$, the BCC phase gains true thermodynamic
stability in an interval of pressures.
Obviously, this careful monitoring of the relative stability of the various
phases as a function of both temperature and pressure would simply be
impossible without the knowledge of their respective chemical potentials,
and this is the ultimate reason for preferring exact free-energy calculations
to other methods.

As far as the experiment is concerned, it is likely that the cusp-like
feature on the Xe freezing line is just an experimental artefact, which
could be due to the increasing difficulty, as the freezing density
progressively grows, in establishing hydrostatic conditions,
a problem also worsened by the existence of two different solid
phases that so closely compete with each other for stability.

\newpage
%
%

\newpage
%
%
\begin{center}
\large
TABLE CAPTIONS
\normalsize
\end{center}
\begin{description}
\item[{\bf Table 1 :}] Excess free energy per particle in
units of $k_BT$ for a number of exp-6 solid states.
For each state and solid type, also shown within square brackets is the
value of $c^*=cr_m^2/\epsilon$ that intervenes the Frenkel-Ladd calculation:
for the chosen $c^*$, $\delta R^2_{\rm Ein}$ approximately matches the average
square deviation of an exp-6 particle from its position in the perfect crystal.

\item[{\bf Table 2 :}] Phase-transition data for $T^*=4.25,8.15,12.77$, and 16.
With the exception of $T^*=4.25$, simulation data refer to $N=1372$
(fluid and FCC) and $N=1458$ (BCC).
For $T^*=4.25$, $N=864$ for fluid and FCC, no BCC phase was considered.
From left to right, values of $\beta P_{\rm SS}$ (BCC-to-FCC ``virtual''
transition, falling inside the fluid region of the phase diagram),
$\beta P_{\rm FS}$ (fluid-to-FCC transition), $\beta\mu_{\rm FS}$
(common value of $\beta\mu$ for the coexisting fluid and FCC solid),
$\beta\Delta\mu_{\rm FS}$ (reduced chemical potential of FCC relative
to BCC at $P=P_{\rm FS}$), $\rho_{\rm f}$ (freezing density), and
$\rho_{\rm m}$ (FCC melting density) are shown.
These quantities were arbitrarily round-off at the third decimal digit
(the fourth digit only for $\beta\Delta\mu_{\rm FS}$).
However, the {\em error} accompanying them would usually be larger,
originating from the limited precision of both the Monte Carlo data
{\em and} the free-energy values in Table 1.

\item[{\bf Table 3 :}] Phase-transition data for $T^*=20$ and 25.
From left to right, values of $\beta P_{\rm FS}$ (fluid-to-BCC transition),
$\beta P_{\rm SS}$ (BCC-to-FCC transition), $\rho_{\rm f}$ (freezing density),
$\rho_{\rm m}$ (BCC melting density), $\rho_{\rm BCC}$ (BCC density at the
BCC-FCC transition), and $\rho_{\rm FCC}$ (FCC density at the BCC-FCC
transition) are shown.
\end{description}

\newpage
%
%
\begin{center}
\large
FIGURE CAPTIONS
\normalsize
\end{center}
\begin{description}
\item[{\bf Fig.\,1 :}] Frenkel-Ladd calculation of solid
free energies: FCC solid at $\rho^*=3.5$ and $T^*=8.15$.
Top: the integrand of Eq.\,(\ref{2-10}) is plotted in the panel above
for $N=1372$ (the continuous line is a spline interpolant of the data
points and the error bars, which are also shown, are much smaller than
the symbols size).
In the panel below, the finite-size effect is demonstrated for
$\beta\left<\Delta V\right>_\lambda/N$ through the difference between
$N=500$ and $N=1372$ ($\triangle$) and between $N=864$ and $N=1372$
($\bigcirc$).
Bottom: the values of $\beta f_{\rm ex}(N)+\ln N/N$ (for $N=500,864$, and
1372) scale linearly with $N^{-1}$ for large $N$, as been conjectured
in Ref.\,\cite{Polson}.
We have verified that the same type of scaling holds for all of the
solid-state points in Table 1.

\item[{\bf Fig.\,2 :}] Reduced chemical potential of the
FCC solid relative to BCC and to HCP for $T^*=16$.
Upon plotting the difference in $\beta\mu$ between FCC and BCC
($N=1372$ vs. $N=1458$, continuous line; $N=864$ vs. $N=1024$,
dashed line), we find that, as pressure grows, the BCC phase loses
stability to the advantage of FCC.
All the plotted lines are linear-spline interpolants of the data points.
The dotted line is the quantity
$\beta(\mu_{\rm FCC}-\mu_{\rm HCP})$ ($N=1372$ for the FCC solid;
$N=1440$ for the HCP solid).
The two couples of vertical lines mark, from left to right, the position
of the ``virtual'' transition between BCC and FCC and the fluid-FCC
phase transition.

\item[{\bf Fig.\,3 :}] Mechanical equation of state for
$T^*=16$. The fluid branch ($\bigcirc$ and continuous line)
and the FCC branch ($\Box$ and dotted line) are plotted for $N=1372$
particles, while the BCC branch ($\triangle$ and dashed line) is for
a system of 1458 particles.
In the inset, we zoom into the transition region: the freezing and
melting densities (signalled by vertical lines) are located where
the horizontal line at $\beta P_{\rm FS}$ crosses the fluid and FCC
equations of state.

\item[{\bf Fig.\,4 :}] Reduced chemical potential of the
FCC ($N=1372$) phase relative to BCC ($N=1458$, continuous line) and HCP
($N=1440$, dotted line), and of the fluid phase ($N=1372$) relative to
BCC ($N=1458$, dashed line), for $T^*=25$.
Upon increasing pressure, we observe a fluid-BCC-FCC sequence of phases.

\item[{\bf Fig.\,5 :}] HT/HP phase diagram of the exp-6 model
(temperature and pressure are rescaled using Ross parameters for Xe).
Besides our data (full and open dots), we plot the coexistence loci
for the exp-6 model as been calculated by Belonoshko {\it et al.}
(dotted lines), and the Xe freezing line as been derived -- after
suitable rescaling of temperature and pressure values -- from the
Simon-Glatzel fit of Ne data by Vos {\it et al.} (dashed line).

\item[{\bf Fig.\,6 :}] Volume-temperature phase diagram
of Xe as drawn from our numerical simulation of the exp-6 model.
Present data for the melting and the freezing volumes ($\Box$ and
continuous lines) and for the BCC-FCC coexistence volumes ($\triangle$
and straight lines) are compared with the estimates by Belonoshko
{\it et al.} ($\times$ and dotted lines, as got from Fig.\,11
of Ref.\,\cite{Belonoshko1}).
Following Fig.\,11 of the cited reference, we have also plotted as
circles some isobars (open circles, fluid; full dots, FCC solid).
\end{description}

\newpage
%
%
\begin{center}
\large
TABLE 1
\normalsize
\vspace{10mm}

\begin{tabular*}{\textwidth}[c]{@{\extracolsep{\fill}}|r||r|r|r|}
\hline
$\beta f_{\rm ex}$ & FCC\,\,\,($N=1372$) & HCP\,\,\,($N=1440$) & BCC\,\,\,($N=1458$) \\
\hline\hline
$\rho^*=4\,,T^*=4.25$ & 38.879(1)\,\,\,$\left[8500\right]$ & & 39.446(2)\,\,\,$\left[5100\right]$ \\
\hline
$\rho^*=3.5\,,T^*=8.15$ & 16.258(1)\,\,\,$\left[5400\right]$ & & 16.408(2)\,\,\,$\left[3000\right]$ \\
\hline
$\rho^*=4\,,T^*=12.77$ & 17.216(1)\,\,\,$\left[7700\right]$ & & 17.316(2)\,\,\,$\left[5300\right]$ \\
\hline
$\rho^*=5\,,T^*=16$ & 25.654(1)\,\,\,$\left[15000\right]$ & 25.659(1)\,\,\,$\left[14500\right]$ & 25.784(2)\,\,\,$\left[9100\right]$ \\
\hline
$\rho^*=5.5\,,T^*=20$ & 27.289(1)\,\,\,$\left[19600\right]$ & 27.295(1)\,\,\,$\left[19400\right]$ & 27.397(2)\,\,\,$\left[13600\right]$ \\
\hline
$\rho^*=5\,,T^*=25$ & 18.489(1)\,\,\,$\left[13800\right]$ & 18.494(1)\,\,\,$\left[14300\right]$ & 18.532(2)\,\,\,$\left[10100\right]$ \\
\hline
\end{tabular*}
\end{center}

\newpage
%
%
\begin{center}
\large
TABLE 2
\normalsize
\vspace{10mm}

\begin{tabular*}{\textwidth}[c]{@{\extracolsep{\fill}}|r||r|r|r|r|r|r|}
\hline
$ T^* $ & $\beta P_{\rm SS}\,\,\,(r_m^{-3})$ & $\beta P_{\rm FS}\,\,\,(r_m^{-3})$ & $\beta\mu_{\rm FS}$ & $\beta\Delta\mu_{\rm FS}$ & $\rho_{\rm f}\,\,\,(r_m^{-3})$ & $\rho_{\rm m}\,\,\,(r_m^{-3})$ \\
\hline\hline
4.25 & -- & 28.897 & 18.774 & -- & 1.980 & 2.055 \\
\hline
8.15 & 40.979 & 46.883 & 25.565 & $-0.0124$ & 2.496 & 2.571 \\
\hline
12.77 & 57.445 & 62.238 & 29.519 & $-0.0060$ & 2.946 & 3.022 \\
\hline
16 & 69.130 & 71.877 & 31.552 & $-0.0029$ & 3.219 & 3.295 \\
\hline
\end{tabular*}
\end{center}

\newpage
%
%
\begin{center}
\large
TABLE 3
\normalsize
\vspace{10mm}

\begin{tabular*}{\textwidth}[c]{@{\extracolsep{\fill}}|r||r|r|r|r|r|r|}
\hline
$ T^* $ & $\beta P_{\rm FS}\,\,\,(r_m^{-3})$ & $\beta P_{\rm SS}\,\,\,(r_m^{-3})$ & $\rho_{\rm f}\,\,\,(r_m^{-3})$ & $\rho_{\rm m}\,\,\,(r_m^{-3})$ & $\rho_{\rm BCC}\,\,\,(r_m^{-3})$ & $\rho_{\rm FCC}\,\,\,(r_m^{-3})$ \\
\hline\hline
20 & 83.517 & 83.575 & 3.533 & 3.601 & 3.602 & 3.611 \\
\hline
25 & 96.901 & 103.84 & 3.888 & 3.958 & 4.039 & 4.048 \\
\hline
\end{tabular*}
\end{center}


\begin{thebibliography}{99}
\bibitem{note}  This follows from a straightforward application of
the rules of functional calculus: if we call $I[f]$ the functional
$\int_a^b{\rm d}x\,f(x)$, the first variation of $I$ is given by
$\delta I[f]=\int_a^b{\rm d}x\,\delta f(x)$.
\end{thebibliography}
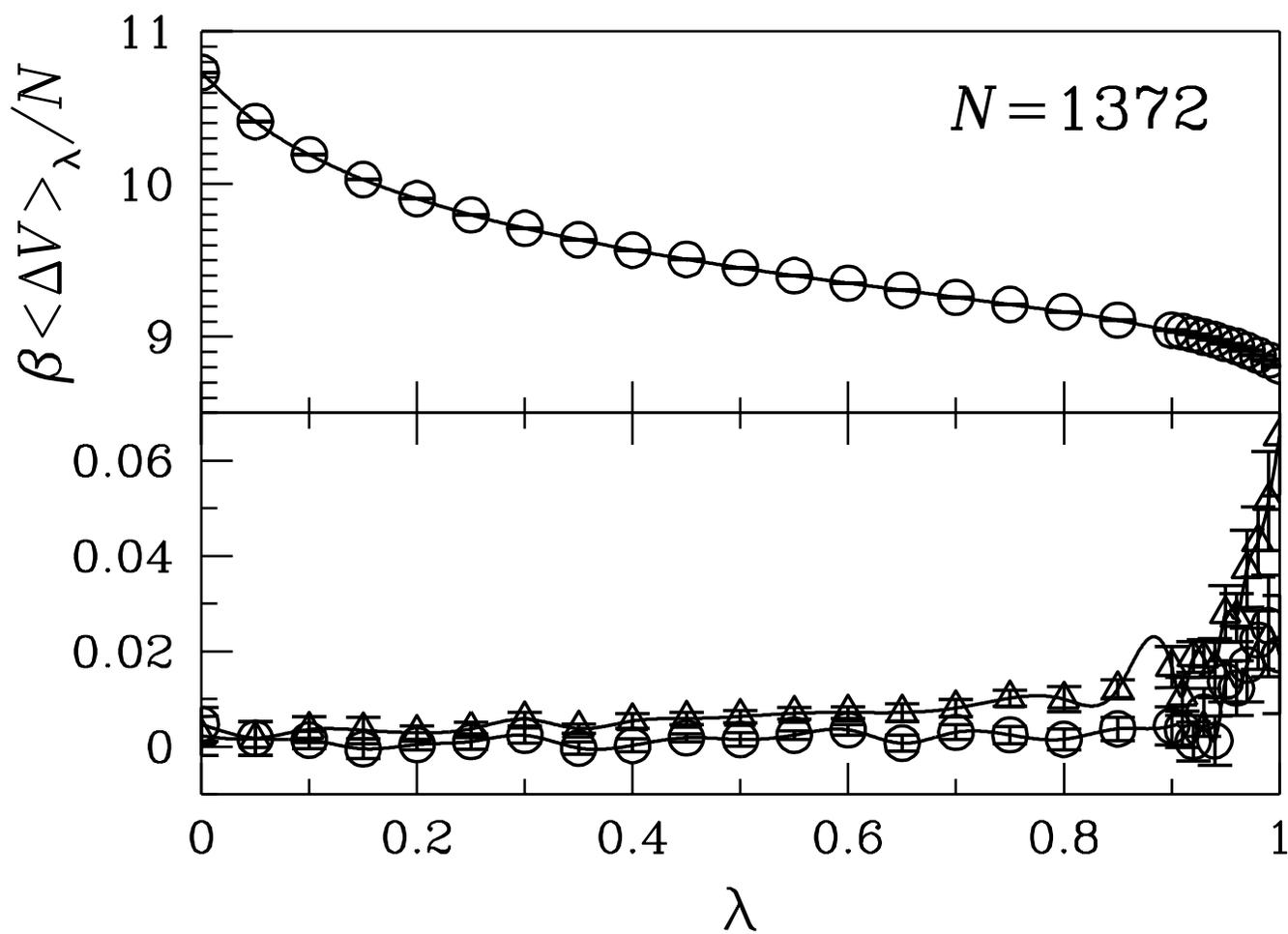
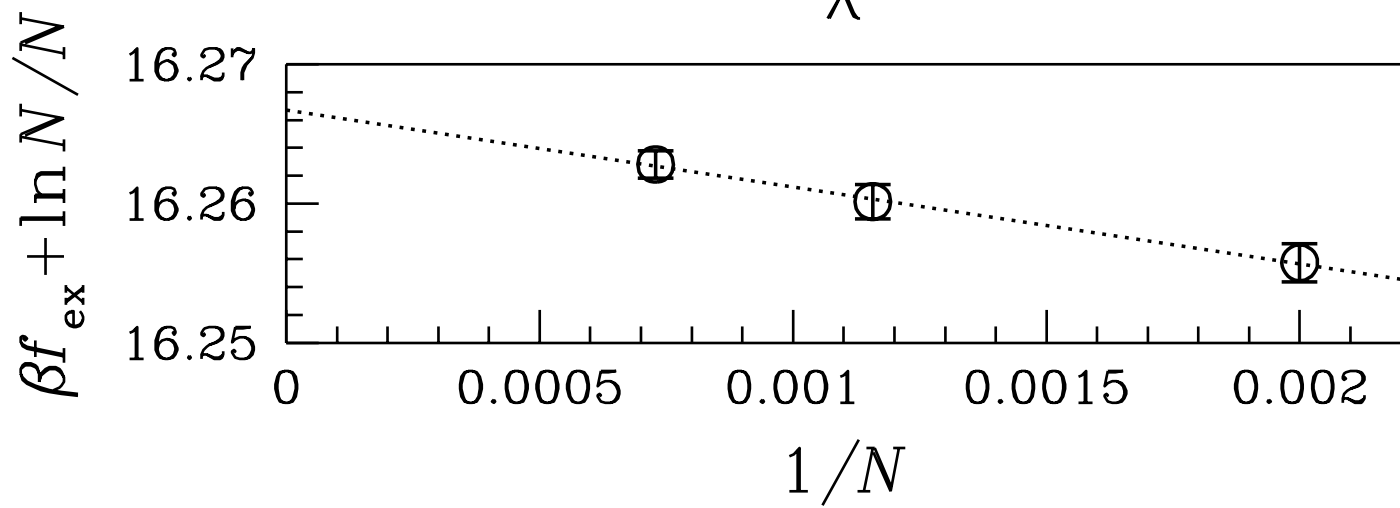
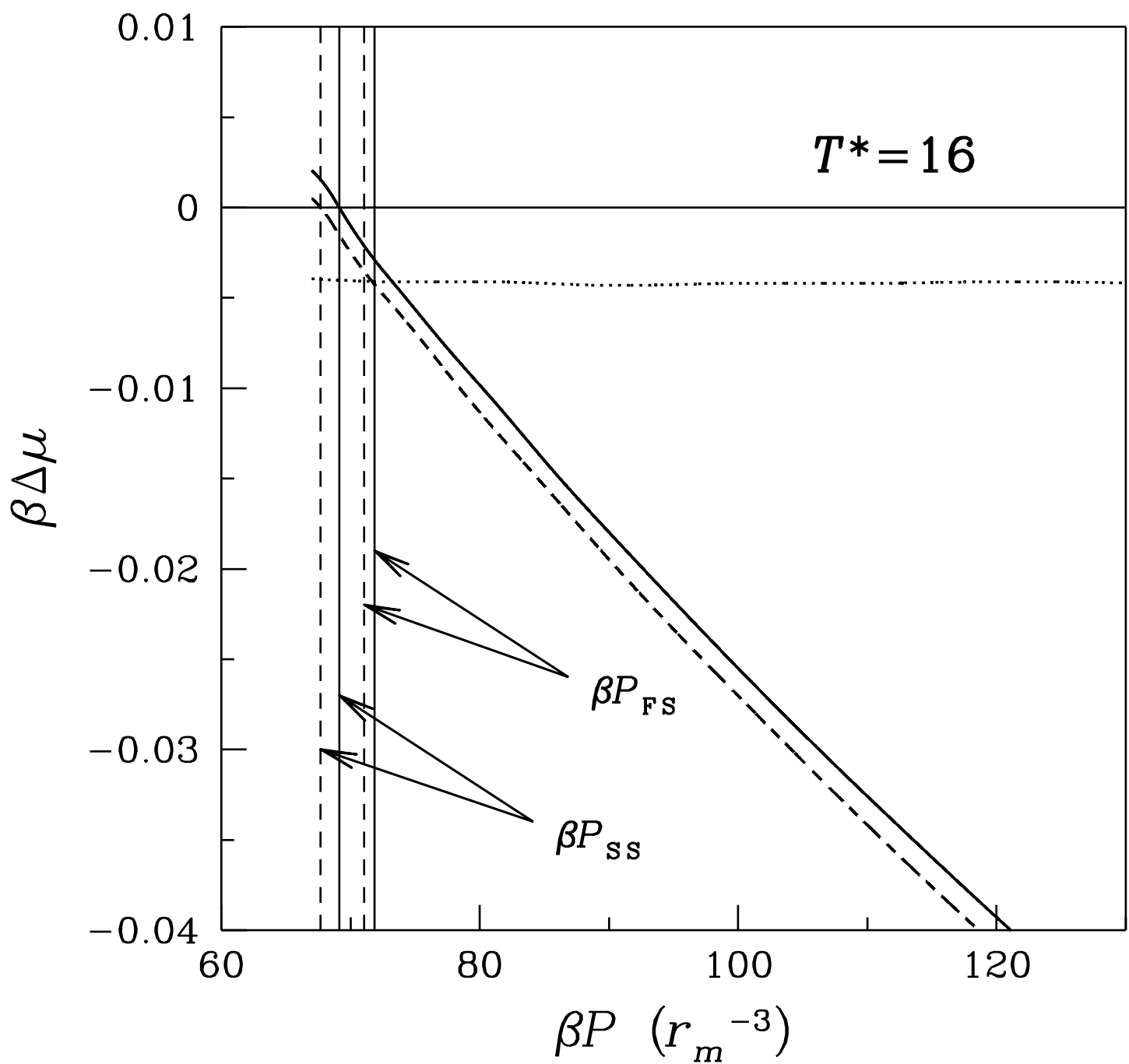
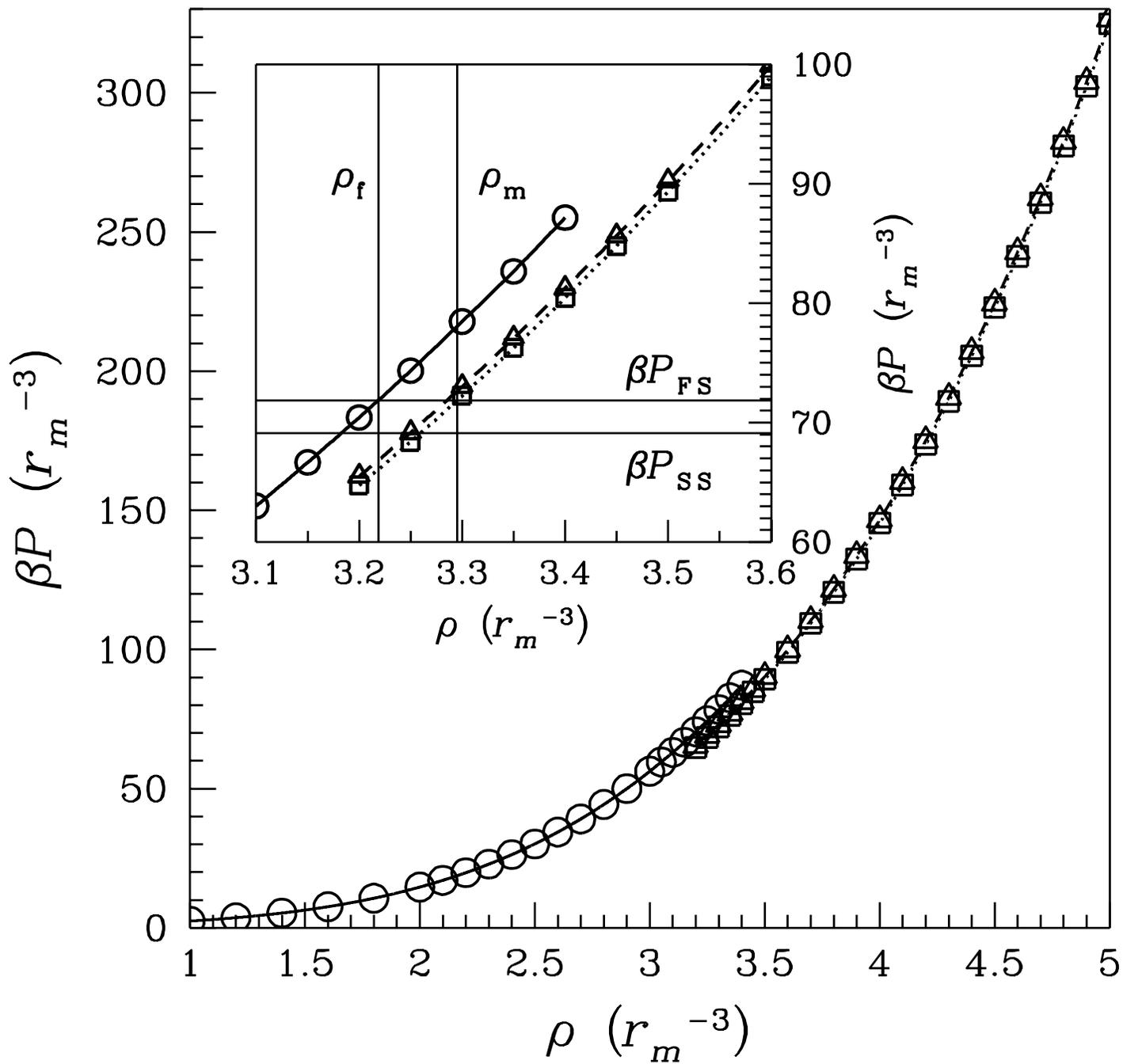
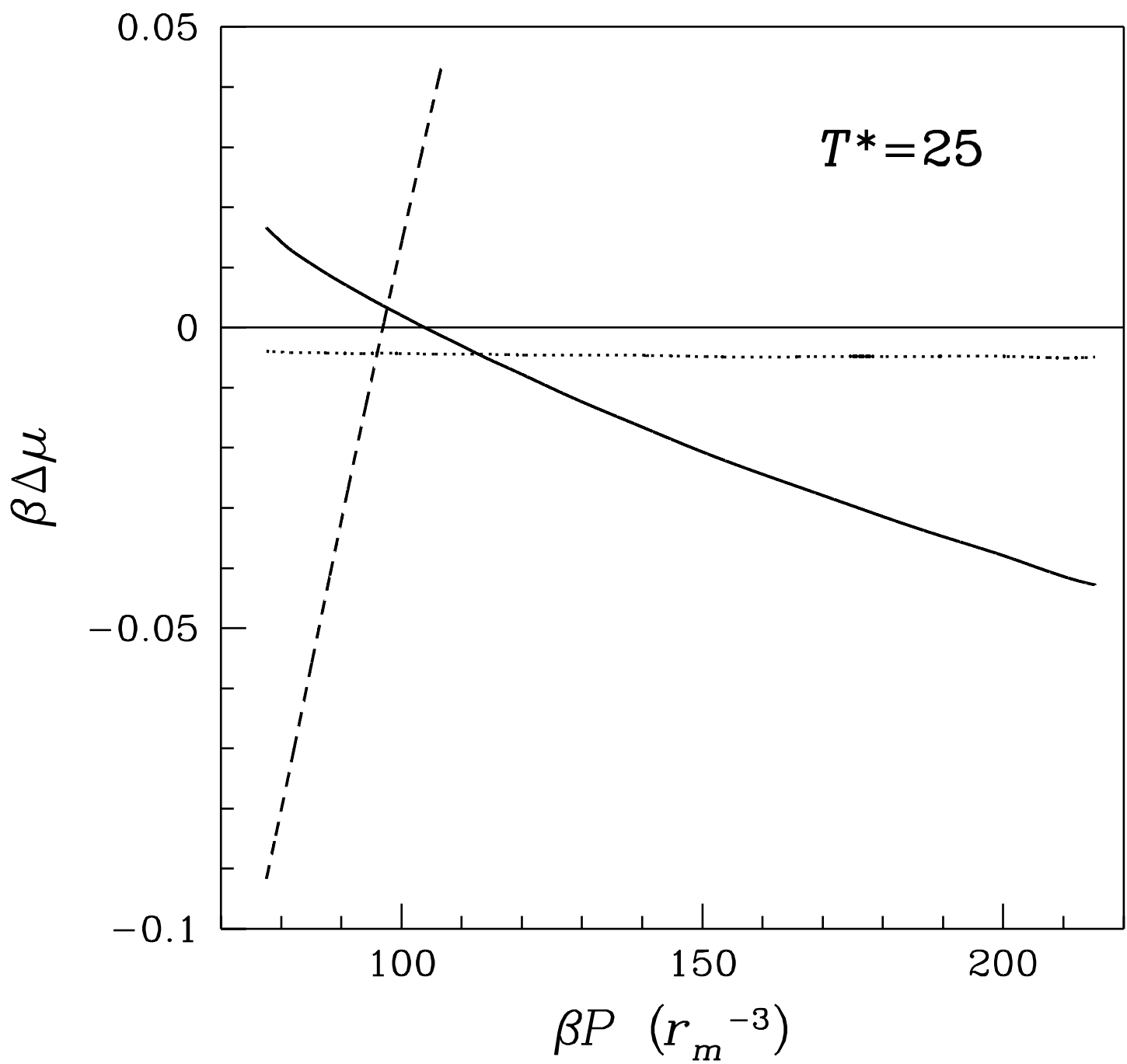
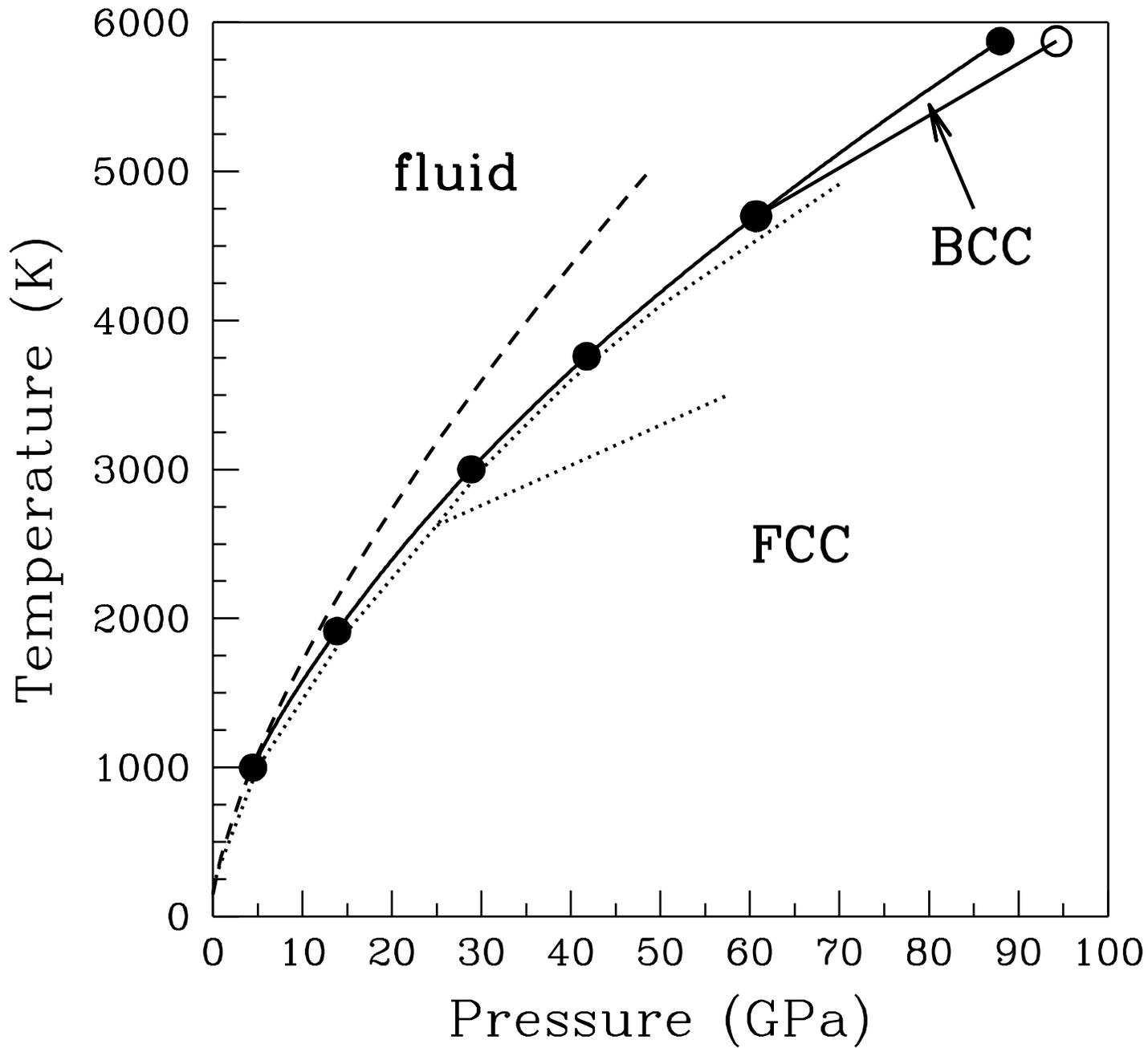
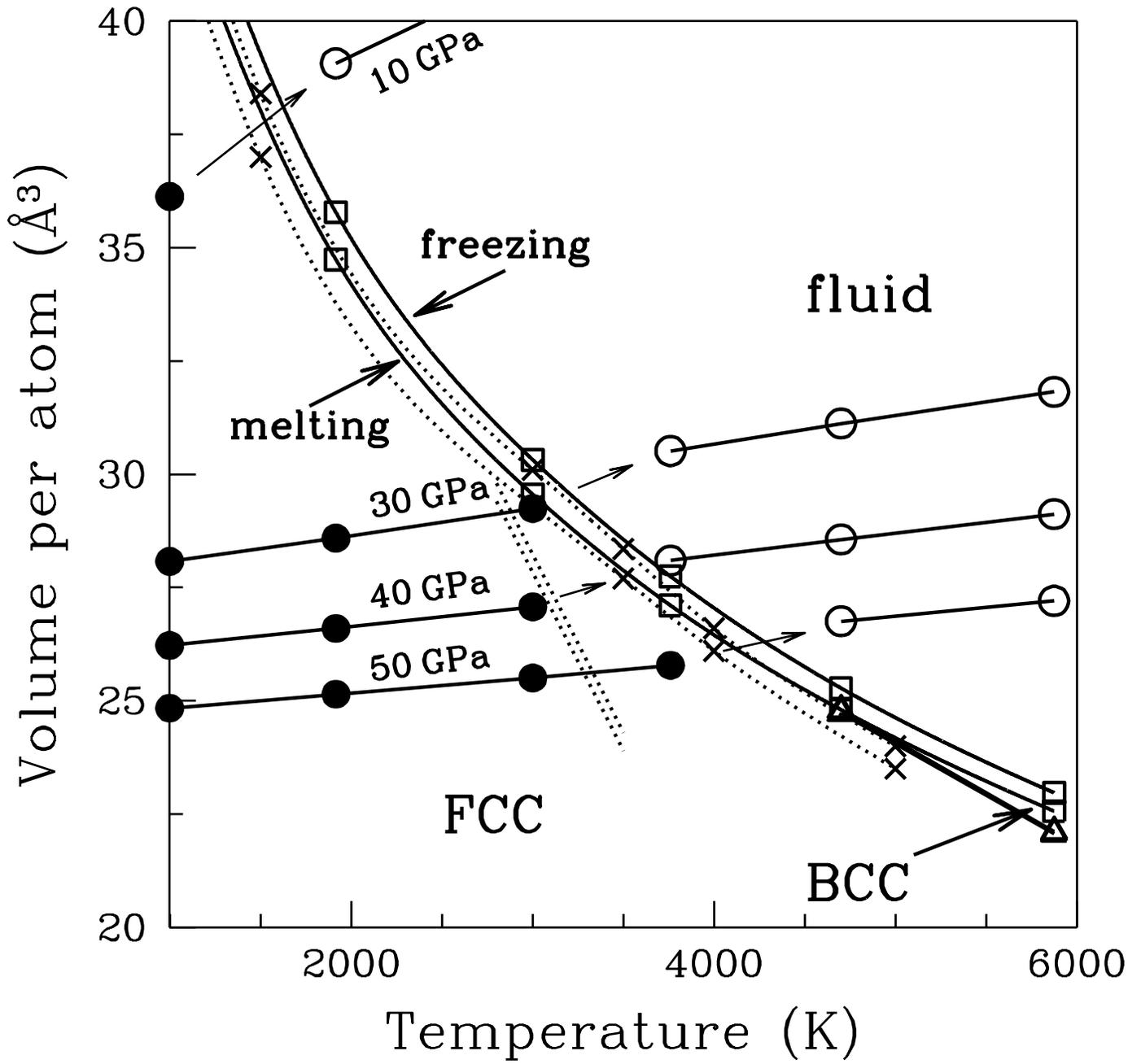
\end{document}